\documentclass{ws-ijgmmp}
\usepackage[figuresright]{rotating}

\begin{document}

\markboth{Marek Szyd{\l}owski, Aleksandra Kurek}
{Towards classification of simple dark energy cosmological models}
\catchline{}{}{}{}{}

\title{Towards classification of simple dark energy cosmological models}

\author{Marek Szyd{\l}owski}

\address{Astronomical Observatory, Jagiellonian University, 
Orla 171, 30-244 Krak{\'o}w, Poland}
\address{Marc Kac Complex Systems Research Centre, Jagiellonian University, 
Reymonta 4, 30-059 Krak{\'o}w, Poland}

\author{Aleksandra Kurek}

\address{Astronomical Observatory, Jagiellonian University, 
Orla 171, 30-244 Krak{\'o}w, Poland}
\maketitle

\begin{abstract}
We characterize a class of simple FRW models filled by both dark energy and dark matter in notion of a single potential function of the scale factor $a(t)$; $t$ is the cosmological time. It is representing potential of fictitious particle - Universe moving in 1-dimensional well $V(a)$ which the positional variable mimics the evolution of the Universe. Then the class of all dark energy models (called a multiverse) can be regarded as a Banach space naturally equipment in the structure of the Sobolev metric. In this paper we explore notion of $C^{1}$ metric introduced in the multiverse which measure distance between any two dark energy models. If we choose cold dark matter as a reference one then we can find how so far apart are different models offering explanation of present accelerating expansion phase of the Universe. We consider both models with dark energy (models with the generalized Chaplygin gas, models with variable coefficient equation of state $w_{X}=\frac{p_{X}}{\rho_{X}}$ parameterized by redshift $z$, models with phantom matter) as well as models basing on some modification of the Friedmann equation (Cardassian models, Dvali-Gabadadze-Porati brane models). We argue that because observational data still favor the $\Lambda$CDM model all reasonable dark energy models should belong to the nearby neighborhood of this model.
\end{abstract}

\section{Introduction -- dark energy models as a dynamical systems of Newtonian type}

Let us assume a class of cosmological models with the Robertson-Walker symmetry (i.e. homogeneous and isotropic) of space slices filled by perfect fluid described by the general form of the equation of state $w_{\mathrm{eff}}=\frac{p_{\mathrm{eff}}}{\rho_{\mathrm{eff}}}=w_{\mathrm{eff}}(a)$, where $p_{\mathrm{eff}}$ and $\rho_{\mathrm{eff}}$ are effective pressure and energy density. Then dynamics of the models under consideration, i.e. the FRW models are governed by two basic equations
\begin{eqnarray}
\frac{\ddot{a}}{a} = - \frac{1}{6} (\rho +3p), \label{eq:1} \\
\dot{\rho}=- 3H(\rho +p), \label{eq:2} 
\end{eqnarray}
where $\rho(a)$ and $p(a)$ plays a role of the effective energy density and pressure respectively, $a$ is the scale factor, $H=(\ln a)\dot{}$ is the Hubble function, a dot means differentiation with respect to the cosmological time $t$; $8 \pi G=c=1 $.

The first equation is called the acceleration equation or Raychaudhuri equation and the second one is the conservation condition. The system of two basic equation will be in closed form after postulating the form of equation of state (E.Q.S). We assume that Universe is filled by mixture of $i$ noninteracting fluids (in general) such that
\begin{equation}\label{eq:3}
p_{\mathrm{eff}}=\sum _{i} p_{i} \equiv \sum_{i} w_{i}(a)\rho_{i}.
\end{equation}
For example if we assume dust mater (baryonic and dark mater) and dark energy $X$ then effective pressure and energy density are function of the scale factor only
\begin{eqnarray}
p_{\mathrm{eff}} &=& 0+w_{X}(a) \rho_{X}(a), \nonumber \\
\rho_{\mathrm{eff}} &=& \rho_{\mathrm{m},0}a^{-3} + \rho_{X,0} a^{-3} \exp \left [ -3 \int_{1}^{a} \frac{w_{X}(a)}{a}da \right ],
\end{eqnarray}
where $w_{X}(a) \equiv \frac{p_{X}}{\rho_{X}}$ is a coefficient of E.Q.S for dark energy, $1+z=a^{-1}$ and $a=1$ is corresponding to a present value of the scale factor.

Hence for assumed form of E.Q.S equation (\ref{eq:1}) can be written down in the form analogous to the Newtonian equation of motion
\begin{equation}\label{eq:5}
\ddot{a}=-\frac{\partial V}{\partial a},\quad V\equiv - \frac{\rho_{eff}a^{2}}{6}.
\end{equation}
In equation \ref{eq:5} $V(a)$ plays the role of the potential function of the particle of unit mass moving in the 1-dimensional potential well. Equation (\ref{eq:5}) reduces problem of dynamics cosmological models with dark energy to the problem of classical mechanics if only energy density satisfies conservation condition (\ref{eq:2}). 

Of course system (\ref{eq:5}) admits the first integral in the form
\begin{equation}\label{eq:6}
\frac{\dot{a}^{2}}{2} + V(a) = E = \mathrm{const},
\end{equation}
which is usually called the Friedmann first integral.

It is convenient to rewrite equation (\ref{eq:6}) to the new (equivalent) form in which appears dimensionless density parameters $\Omega_{i}=\frac{\rho_{i}}{3H_{0}^{2}}$. For this aims it is sufficient to reparameterize the cosmological time: $t \mapsto \tau \colon |H_{0}|dt=d\tau$. Then we obtain
\begin{equation}\label{eq:7}
\frac{y^{2}}{2} + V(x) = \frac{1}{2} \Omega_{k,0},
\end{equation}
where $x \equiv \frac{a}{a_{0}}$, $y=x'$, ${}' \equiv \frac{d}{d\tau}$ ($a_{0}$ is present value of the scale factor), $\Omega_{k}=\Omega_{k,0}a^{-2} \equiv \frac{-3 \frac{k}{a^{2}} }{3H_{0}^{2}}$, $1+z=\frac{a_{0}}{a}=\frac{1}{a}$. Without lost degree of generality, $k$ (curvature index) and density parameter for curvature can be formally incorporated into potential function.

Therefore the motion of the system is restricted to the distinguished energy level $E=\frac{1}{2} \Omega_{k,0}$. In the phase ($x,y$) space trajectories of the system are determined from relation (\ref{eq:7}). Therefore the classical motion of the system in configuration space is restricted to the domain admissible for motion
$$
\mathcal{D}_{E} = \{ x \colon E-V(x) \ge 0 \}.
$$
In the special case of the $\Lambda$CDM model ($\Omega_{k,0}=0$) we have
\begin{eqnarray}
V(x)=- \frac{1}{2} \{\Omega_{\mathrm{m},0}x^{-1} + \Omega_{\Lambda,0}x^{2} \}, \label{eq:8}\\
V(x(z))=- \frac{1}{2} \{\Omega_{\mathrm{m},0}(1+z) + (1-\Omega_{\mathrm{m},0})(1+z)^{-2} \},  
\nonumber
\end{eqnarray}
where we use constraint relation $\Omega_{m,0}+ \Omega_{\Lambda,0}=1$.

Formally curvature term can be included into potential but then system should be considered on zero energy $E=0$ level instead of $E=\frac{1}{2}\Omega_{k,0}$. If we include the curvature term into the potential $V$ then an additional additive term appears.

The phase space ($x,y=x'$) is divided by trajectory of the flat model ($\Omega_{k,0}=0$) $y^{2}=-2V(x)$ on domains occupied by close ($\Omega_{k,0} < 0$) and open models ($\Omega_{k,0} > 0$).

In the general case because first integral is valid for any $z$ we obtain constraint condition $H(z=0)=H_{0}$ in the form
\begin{equation}\label{eq:9}
\Omega_{\mathrm{m},0} + \Omega_{X,0} + \Omega_{k,0}=1.
\end{equation}
We complete the different FRW models with dark energy (see Table \ref{tab:1}) and different generalized FRW models based on modification of the Friedmann equation (see Table \ref{tab:2}), appeared in the context of discussion in the origin of the present accelerating phase of the Universe \cite{Szydlowski:2006pz,Szydlowski:2006ay}. For all these models the dynamics is determined from the potential function $V(x(z))$ and corresponding dynamical system is of the form
\begin{equation}\label{eq:10}
x' = y;\quad y' = - \frac{\partial V(x) }{\partial x}; \quad
\mathrm{where} \ \frac{y^{2}}{2} + V(x) \equiv \frac{1}{2} \Omega_{k,0} 
\end{equation}
i.e. a Newtonian type. Therefore in the phase space (at finite domain) only critical points of the saddle or centre types are admissible. It is simple consequence of the fact that characteristic equation of the linearization matrix $\lambda^{2} + V_{xx}(x_{0})=0$ calculated at the critical points $(y_{0}=0,\ V(x_{0})=\frac{1}{2}\Omega_{k,0})$ are admissible. In any case they are corresponding to extremum of the potential.  If $V_{x}(x_{0})=0$ and $V_{xx}(x_{0})<0$ then we obtain saddle points (eigenvalues are real of opposite signs). If at the critical point $V_{xx}(x_{0}) >0 $ then we obtain centre.

From the physical point of view max is corresponding moment of switching decelerate phase dominate by matter on accelerate dominating by dark energy effects.
\begin{sidewaystable}
\caption{The potential function for 10 prototypes of cosmological models 
explaining the present acceleration of the Universe in terms of dark energy}
\label{tab:1}
\begin{tabular}{p{0.5cm}p{4cm}p{0.3cm}l}
\hline
case & model && $V(x(z))$ relation \\
\hline 
&&&\\
1&\small $\Lambda$CDM model& &
 \small $  V(x(z))=-\frac{1}{2}\{\Omega_{\mathrm{m},0}(1+z)+(1- \Omega_{\mathrm{m},0})(1+z)^{-2}\}$ \smallskip  \\
2& \small non-flat FRW model with $\Lambda$& &
 \small $V(x(z))=-\frac{1}{2} \{\Omega_{\mathrm{m},0}(1+z)+\Omega_{k,0} + 
 \Omega_{\Lambda,0}(1+z)^{-2}\}$ \smallskip
  \\
3&\small FRW model with 2D topological defects $p_{X}=-\frac{2}{3} \rho_{X}$& &
 \small $V(x(z))=-\frac{1}{2}\{\Omega_{\mathrm{m},0}(1+z)+\Omega_{k,0} + 
 \Omega_{\mathrm{top,0}}(1+z)^{-1}\}$ \smallskip
 \\
4 &\small FRW model with phantom dark energy $p_{X}=-\frac{4}{3} \rho_{X}$& &
 \small $V(x(z))=-\frac{1}{2}\{\Omega_{\mathrm{m},0}(1+z)+\Omega_{k,0} + 
 \Omega_{ph,0}(1+z)^{-3}\}$ \smallskip 
  \\
5&\small FRW model with phantom dark energy $p_{X}=w_{X}\rho_{X}$, $w_{X}<-1$ fitted&& 
 \small $V(x(z))=-\frac{1}{2}\{\Omega_{\mathrm{m},0}(1+z)+\Omega_{k,0} + 
 \Omega_{\mathrm{ph},0}(1+z)^{3(1+w_{X})-2}\}$ \smallskip
  \\
6 &\small FRW model with Chaplygin gas $p_{X}=-\frac{A}{\rho_{X}},\ A>0$&&
 \small $V(x(z))=-\frac{1}{2} \left\{\Omega_{\mathrm{m},0}(1+z)+\Omega_{k,0} + 
 \Omega_{\mathrm{Ch},0}(1+z)^{-2}[A_{S}+(1-A_{S})(1+z)^{6}]^{ \frac{1}{2}} \right\}$ \smallskip
 \\
7&\small FRW model with generalized Chaplygin gas 
 $p_{X}=-\frac{A}{\rho_{X}^{\alpha}},$ $A>0$, $\alpha = \mathrm{const}$&&
\small $V(x(z))= -\frac{1}{2}\left\{\Omega_{\mathrm{m},0}(1+z)+
 \Omega_{k,0} + \Omega_{\mathrm{Ch},0}(1+z)^{-2}[A_{S}+(1-A_{S})
 (1+z)^{3(1+\alpha)}]^{\frac{1}{1+\alpha}}\right\}$ \smallskip
 \\
8a &\small FRW models with dynamical E.Q.S. parameterized by $z$ 
 $p_{X}=(w_{0}+w_{1}z)\rho_{X}$&&
\small $V(x(z))=-\frac{1}{2}\left\{\Omega_{\mathrm{m},0}(1+z)+ \Omega_{k,0} + 
 \Omega_{X,0}(1+z)^{3(w_{0}-w_{1}+1)-2} \exp [3w_{1}z] \right\}$ \smallskip 
  \\
8b &\small FRW models with dynamical E.Q.S parameterized by scale factor $a$
 $p_{X}=(w_{0}+$ $w_{1}(1-a))\rho_{X}$&&
 \small $V(x(z))=-\frac{1}{2} \left\{\Omega_{\mathrm{m},0}(1+z)+\Omega_{k,0} + 
 \Omega_{X,0}(1+z)^{3(w_{0}+w_{1}+1)-2} \exp [- \frac {3w_{1}z}{1+z}] \right\}$\smallskip 
 \\
9&\small FRW model with quantum effects origin from massless scalar field at 
 low temperature (Casimir effect) $\rho_{X}=-\frac{\rho_{X,0}}{a^{4}},$ 
 $\rho_{X,0}>0 $&&
 \small $V(x(z))=-\frac{1}{2}\{\Omega_{\mathrm{m},0}(1+z)+\Omega_{k,0} + 
 \Omega_{\Lambda,0}(1+z)^{-2} - \Omega_{\mathrm{Cass},0}(1+z)^{2}\}$ \smallskip
  \\
10&\small flat FRW model with dissipative dark energy 
 $p_{\mathrm{eff}}=0-3 \bar{\alpha} \rho^{m}H$; 
 $m=-1.5$ is fixed &&
 \small $V(x(z))=-\frac{1}{2} \left \{ \Omega_{m,0}(1+z) + \Omega_{\mathrm{diss},0}(1+z)^{-2}[A_{S}+
 (1-A_{S}) (1+z)^{3(\frac{1}{2}-m)}]^{\frac{2}{1-2m}} \right\}$ \smallskip
  \\
\hline
\end{tabular}
\end{sidewaystable}

\begin{sidewaystable}
\caption{The potential function for 10 cosmological models beyond 
the standard general relativity} 
\label{tab:2}
\begin{tabular}{p{0.5cm}p{4cm}p{0.3cm}l}
\hline
 case & model && $V(x(z))$ relation \\
\hline 
 & &&\\
1&\tiny Cardassian type of Friedmann equation, $\Omega_{r,0}=10^{-4}$ is fixed && 
\tiny $V(x(z))=-\frac{1}{2} \left \{ \Omega_{k,0}+\Omega_{\mathrm{m},0}(1+z)^{2} 
 \left[ \frac{1}{1+z} + (1+z)^{-4+4n} \left ( \frac{1- \Omega_{r,0}-
 \Omega_{\mathrm{m},0}}{\Omega_{\mathrm{m},0}} \right ) \left ( \frac{ \frac{1}{1+z}+
 \frac{\Omega_{r,0}}{\Omega_{\mathrm{m},0} }}{ 1+ 
 \frac{\Omega_{r,0}}{\Omega{m,0}}}\right)^{n} \right ] \right \} $
  \smallskip \\
2&\tiny Dvali-Gabadadze-Porrati brane models (DGP) &&
 \tiny $V(x(z))=-\frac{1}{2} \left \{ (1+z)^{-2} \left [ \sqrt{\Omega_{\mathrm{m},0}(1+z)^{3}+\Omega_{rc,0}}+
 \sqrt{\Omega_{rc,0}} \right] ^{2} + \Omega_{k,0} \right \}$ 
 \smallskip \\
3&\tiny Deffayet-Dvali-Gabadadze brane models with $\lambda$ (DDG) &&
 \tiny $V(x(z))=-\frac{1}{2} \left \{ (1+z)^{-2}\left [ - \frac{1}{2r_{0}H_{0}} +
 \sqrt{\Omega_{\mathrm{m},0}(1+z)^{3}+\Omega_{\lambda,0} + 
 \frac{1}{4r_{0}^{2}H_{0}^{2}}}\right]^{2} \right \}$ 
  \smallskip \\
4&\tiny Randall-Sundrum brane models with dark radiation and $\Lambda=0$ &&
 \tiny $V(x(z))=-\frac{1}{2} \left \{ \Omega_{\mathrm{m},0}(1+z)+\Omega_{k,0}+ 
 \Omega_{dr,0}(1+z)^{2}+ \Omega_{\lambda,0}(1+z)^{4} \right \}$ 
 \smallskip \\
5&\tiny Randall-Sundrum brane models with dark radiation and $\Lambda$ (RSB) &&
 \tiny $V(x(z))=-\frac{1}{2} \left \{ \Omega_{\mathrm{m},0}(1+z)+\Omega_{k,0}+ 
 \Omega_{dr,0}(1+z)^{2}+ \Omega_{\lambda,0}(1+z)^{4} + \Omega_{\Lambda,0}(1+z)^{-2} \right \}$ 
 \smallskip \\
6a&\tiny Shtanov brane models (Brane1) &&
 \tiny $V(x(z))=-\frac{1}{2} \left \{ \Omega_{\mathrm{m},0}(1+z)+\Omega_{\sigma,0}(1+z)^{-2}+ 
 2 \Omega_{l,0}(1+z)^{-2}-2(1+z)^{-2}\sqrt{\Omega_{l,0}}\sqrt{\Omega_{\mathrm{m},0}(1+z)^{3}+
 \Omega_{\sigma,0}+ \Omega_{l,0}+ \Omega_{\Lambda b,0}} \right \}$ 
 \smallskip \\
6b.&\tiny Shtanov brane models (Brane2)&& 
 \tiny $V(x(z))=-\frac{1}{2} \left \{ \Omega_{\mathrm{m},0}(1+z)+\Omega_{\sigma,0}(1+z)^{-2}+ 
 2 \Omega_{l,0}(1+z)^{-2}+2(1+z)^{-2}\sqrt{\Omega_{l,0}}\sqrt{\Omega_{\mathrm{m},0}(1+z)^{3}+
 \Omega_{\sigma,0}+ \Omega_{l,0}+ \Omega_{\Lambda b,0}}\right \}$
 \smallskip \\
7&\tiny modified affine gravity (MAG) model&&
 \tiny $V(x(z))=-\frac{1}{2} \left \{ \Omega_{\mathrm{m},0}(1+z)+\Omega_{k,0}+ 
 \Omega_{\psi,0}(1+z)^{4}+ \Omega_{\Lambda,0}(1+z)^{-2} \right \}$
 \smallskip \\
8&\tiny FRW models of nonlinear gravity with Lagrangian density proportional to
 Ricci scalar $R$ (NG) &&
 \tiny $V(x(z))=-\frac{1}{2} \left \{ \left \{\Omega_{\mathrm{m},0}(1+z)\frac{2n}{3-n}+
 \Omega_{r,0}(1+z)^{2} \frac{4n(2-n)}{(n-3)^{2}} \right \}
 \Omega_{\mathrm{nonl},0}(1+z)^{\frac{3(1-n)}{n}-2} \right \}$
  \smallskip \\
9&\tiny bouncing models with $\Lambda$ (B$\Lambda$CDM) &&
 \tiny $V(x(z))=-\frac{1}{2} \left \{ \Omega_{\mathrm{m},0}(1+z)+\Omega_{k,0}- 
 \Omega_{n,0}(1+z)^{n-2}+ \Omega_{\Lambda,0}(1+z)^{-2} \right \}$
 \smallskip \\
10a&\tiny models with energy transfer (dark matter $\leftrightarrow$ vacuum energy 
  $\rho_{1}$ sector)&&
 \tiny $V(x(z))=-\frac{1}{2} \left \{ \Omega_{\mathrm{m},0}(1+z)+\Omega_{\mathrm{int}}(1+z)^{n-2}+ 
  \Omega_{\Lambda,0}(1+z)^{-2}\right \}$
   \smallskip \\
10b&\tiny models with energy transfer (dark matter $\leftrightarrow$ phantom dark 
  energy sector) &&
  \tiny $V(x(z))=-\frac{1}{2} \left \{ \Omega_{\mathrm{m},0}(1+z)+\Omega_{\mathrm{int}}(1+z)^{n-2}+ 
  \Omega_{\mathrm{ph}}(1+z)^{3(1+w_{X})-2}\right \} $ 
  \\
  & &&  \\

\hline
\end{tabular}
\end{sidewaystable}
While potential function $V(a)$ contain all what is needed to characterize the system evolution it is possible to reconstruct this modulo $\Omega_{k,0}$ immediately from the SNIa observations due to the simple relation between the luminosity distance and the Hubble function (for flat models)
\begin{equation}\label{eq:11}
H(z)=\left [ \frac{d}{dz} \left ( \frac{d_{L}(z)}{1+z} \right ) \right ] ^{-1},
\end{equation}
where 
\begin{eqnarray}\label{eq:12}
H^{2}(x(z))= -2 V(x) x ^{-2},\\
1+z=x^{-1}. \nonumber
\end{eqnarray}  
Now we can define notion of space all dark energy cosmological models in terms of space of all dynamical systems of Newtonian type with potential describing dark energy model.

\begin{definition}
By multiverse of cosmological models with dark energy we understood functional space of all dynamical systems of Newtonian type with potential functions $V(x)$ of $C^{1}$ class.
\end{definition}

Introduced before notion of the multiverse of dark energy models satisfying equation of state (\ref{eq:3}) can be generalized on the case of cosmological models with scalar fields. Let us consider for example the FRW model with single minimally coupled to gravity scalar field. They can be defined in the form of perfect fluid with energy density $\rho_{ \phi } = \frac{1}{2} \dot{\phi}^{2} + V(\phi)$ and $p_{ \phi } = \frac{1}{2} \dot{\phi}^{2} - V(\phi)$. Therefore potential function of the system is $U(\phi)=-\frac{1}{12}\dot{\phi}^{2}a^{2} - \frac{1}{6}V(\phi)a^{2}$ and then term  $-\frac{1}{12}\dot{\phi}^{2}a^{2}$ can be shifted into kinetic part because of conventional character of division on kinetic and potential parts for the systems with natural Lagrangian. 
Finally we obtain the dynamical system in the form of energy function
\begin{equation}\label{eq:13}
\mathcal{\epsilon}(a,\dot{a}, \phi, \dot{\phi})= \frac{\dot{a}^{2}}{2} - \frac{1}{12} \dot{\phi}^{2}a^{2} + U(a,\phi),   
\end{equation}
where $U(a,\phi)= -\frac{1}{6} V(\phi) a^{2} - \frac{1}{6} \rho a^{2}$.

The analogous construction can be performed for conformally coupled scalar field. Therefore if we fix the form of kinetic energy then the potential function $U(a, \phi)$ identify quintessence model.

\section{Metric structure multiverse of dark energy cosmological systems}

The space of all dynamical systems on the plane can be naturally equipment in the structure of Banach space after introducing the Sobolev norm within a space of vector field which are tangent to phase curve 
\begin{equation}\label{eq:14}
||f||_{r}=\max_ {x \in C} \left \{ |f(x)|, |\partial_{1}f|,\dots,|\partial_{r}f| \right \},
\end{equation}
where $C$ is closed subspace of the phase space, $f$ is the vector field defined on the phase space; $f=[x, -\frac{\partial V}{\partial x} ]^{T}$ in our case.

Let us consider $C^{1}$ metric defined as $d(f,g)= ||f-g||$. If we consider two models of the multiverse then $C^{0}$ distance between any two dark energy models are
\begin{equation}\label{eq:15}
d_{C^{0}}(1,2) = \max_{x \in C} \left \{ \left | (V_{1})_{x} - (V_{2})_{x} \right | \right \},
\end{equation}
where the curvature term $\frac{\Omega_{k,0}}{2}$ we shift into potential then system is considered on $E=0$ level. Therefore two dark energy models are close if their potential function and first derivatives are close. The choice of subset $C$ of configuration space can be performed in different way. If we are interesting in comparison any two dark energy models from view point of present day observational data. At the present epoch $x=1\ (z=0)$ and $V(1)=-\frac{1}{2} - \frac{1}{2} \Omega_{k,0}$ and (\ref{eq:15}) reduces to
\begin{eqnarray}\label{eq:16}
d(1,2)= \frac{1}{2} \max_{ C=\{1\} x \{ y:\ 0 \le \sqrt{\Omega_{k,0}^{max}-2V(1)}\}} \Bigg \{ \Big | (\Omega_{m,0}^{(1)} - \Omega_{m,0}^{(2)})+ \\
+ \left ( \Omega_{X}^{(1)}(1+3w_{X}^{(1)}(1)) - \Omega_{X}^{(2)}(1+3w_{X}^{(2)}(1) ) \right ) \Big | \Bigg \}, \nonumber
\end{eqnarray}
where we arbitrary assume roughly constraint to $\Omega_{k,0}$ for example that $-0.1 \le \Omega_{k,0} \le 0.1 $, and a set $C$ chosen around $\Omega_{k,0}=0$ on the line $x=1$. Of course if we introduce the $C^{1}$ metric as 
\begin{equation}\label{eq:17}
d_{C^{1}}(1,2)= \max_{(x,y) \in C } \left \{ |(V_{1})_{x}- (V_{2})_{x}|,\  |(V_{1})_{xx} - (V_{2})_{xx}| \right \}
\end{equation}
it will be precisely differentiate between the models. From the observational point of view this metric controls both deceleration and jerk parameters. We say that any two models are close in sense of the $C^{1}$ metric if both deceleration and jerk parameters are close. The $C^{1}$- metric can be naturally generalized on the case of the $C^{r}$ metric which can be expressed in dimensionless parameters controlling all $(r+1)$ time derivatives of the scale factor.

It is convent to rewrite $C^{0}$ metric to the new form
\begin{equation}\label{eq:18}
d_{C^{0}} (1,2) = \max_{C} \left \{ \left | \left [ \frac{d}{dz} \left ( \frac{ H_{(1)}^{2}(z) }{ H_{0}^{2} } \right ) \right ] _{z=0} - \left [ \frac{d}{dz} \left ( \frac{H_{(2)}^{2}(z)}{H_{0}^{2} } \right ) \right]_{z=0} \right | \right \},
\end{equation}
where we apply relation
$$ V_{x}(z=0)= \left [ \frac{ \frac{d}{dz} H^{2}(z) } { H_{0}^{2} } -1 \right ]_{z=0} $$
and index in the parent denotes kind of model taken from Table 1 or Table 2.

The form of the metric (\ref{eq:18}) can be simply generalized on the case of $C^{1}$ metric, namely
\begin{eqnarray}\label{eq:19}
d_{C^{1}}(1,2) = H_{0}^{-2} \max _{(1,y) \in C } \Bigg \{ \left | \left [ \frac{d}{dx} \left ( H_{(1)}^{2}(x) - H_{(2)}^{2}(x) \right ) \right ]_{x=1} \right | ,\\ 
\left | \left [ \frac{d^{2}}{dx^{2}} \left (H_{(1)}^{2}(x) - H_{(2)}^{2}(x) \right ) \right ]_{x=1} + 4 \left [ \frac{d}{dx} \left ( H_{(1)}^{2}(x) - H_{(2)}^{2}(x) \right ) \right ] _{x=1} \right | \Bigg \} . \nonumber
\end{eqnarray}
For our aims it is more suitable the notion of $C^{k}$ norm defined on the class of $C^{k}(E)$ functions, i.e. for $f \in C^{k}(E)$, where $E$ is an open subset of $\mathbb{R}^{n}$ we define
\begin{equation}
||f||_{k} = \sup_{E} |f(x)| + \sup_{E} ||Df(x)|| + \dots + \sup_{E} ||D^{k}f(x)||.
\end{equation}
The norm $||\ ||$ on the right-hand side of this equation is defined in the following way 
$$ ||D^{k}f(x)||= \max \left | \frac{\partial f^{k}(x)}{\partial x_{j1} \dots \partial x_{jk} } \right | $$, the maximum being taken over $j1, \dots, jk = 1, \dots, n$. Each of the spaces $C^{k}(E)$ is then the Banach space \cite{Perko:1996}.

The above metric of relatively character can be useful in comparing to different models with the concordance $\Lambda$CDM one.

Let us illustrate how the Banach metric works for the different dark energy models completed in the Table. Let us consider for simplicity the $C^{0}$ metric. Then we obtain
$$d_{C^{0}}(1,2)=|V_{1x}-V_{2x}|= \left |\Omega_{\mathrm{m},0}^{(1)} + (1-\Omega_{\mathrm{m},0}^{(1)}) -  \Omega_{\mathrm{m},0}^{(2)} - (1-\Omega_{\mathrm{m},0}^{(2)}) -\Omega_{k,0} \right |.$$
Therefore if we take into account independent prior on density parameter for matter $\Omega_{\mathrm{m},0}^{(1)}= \Omega_{\mathrm{m},0}^{(2)}= \Omega_{\mathrm{m},0}^{*}$ from independent galactic observation then we obtain $d_{C^{0}}(1,2)=|\Omega_{k,0}|$. The results of calculation of the $C^{0}$ metric for other models (flat for simplicity) are
$$
\begin{array}{l}
d_{C^{0}}(3,1)=\Omega_{\mathrm{top},0} = (1-\Omega_{\mathrm{m},0}^{*})\\
d_{C^{0}}(4,1)=\Omega_{\mathrm{ph},0}=(1-\Omega_{\mathrm{m},0}^{*})\\
d_{C^{0}}(5,1)=3|1+w_{X}| (1-\Omega_{\mathrm{m},0}^{*})\\
d_{C^{0}}(6,1)=3(1-A_{s}) (1-\Omega_{\mathrm{m},0}^{*})\\
d_{C^{0}}(7,1)= 3(1-A_{s})(1- \Omega_{\mathrm{m},0}^{*}\\
d_{C^{0}}(8a,1)= 3 (1-\Omega_{\mathrm{m},0}^{*}) |w_{0}+1|\\ 
d_{C^{0}}(8b,1)=d(8a,1)\\
d_{C^{0}}(9,1)=4(1-\Omega_{\mathrm{m},0}^{*}-\Omega_{\Lambda,0})\\
d_{C^{0}}(10,1)= 3(1-A_{s}) (1-\Omega_{\mathrm{m},0}^{*})\\
\end{array}
$$
where $w_{X},A_{s},w_{0},\Omega_{\Lambda,0}$ should be taken from estimation with fixed value of $\Omega_{\mathrm{m},0}^{*}$.

The metric $d$ describe residuals with respect to the reference $\Lambda$CDM model controlled by second derivatives of the scale factor. The higher derivatives of the scale factor will be controlled by a more subtle $C^{1}$ metric:
$$d_{C^{1}}(1,2)= |(V_{1})_{x}(x=1)-(V_{2})_{x}(x=1) | + |(V_{1})_{xx}(x=1)-(V_{2})_{xx}(x=1) |=$$
$$=H_{0}^{-2} \Bigg \{  \left | \left [ \frac{d}{dx} \left ( H_{(1)}^{2}(x) - H_{(2)}^{2}(x) \right ) \right ]_{x=1} \right | +$$ 
$$\left | \left [ \frac{d^{2}}{dx^{2}} \left (H_{(1)}^{2}(x) - H_{(2)}^{2}(x) \right ) \right ]_{x=1} + 4 \left [ \frac{d}{dx} \left ( H_{(1)}^{2}(x) - H_{(2)}^{2}(x) \right ) \right ] _{x=1} \right | \Bigg \} $$
The results of calculation $C^{1}$ metric for flat models from Table 1 are
$$
\begin{array}{l}
d_{C^{1}}(2,1)=0\\
d_{C^{1}}(3,1)=3(1-\Omega_{\mathrm{m},0}^{*})\\
d_{C^{1}}(4,1)=5(1-\Omega_{\mathrm{m},0}^{*})\\
d_{C^{1}}(5,1)=3(1-\Omega_{\mathrm{m},0}^{*})|1+w_{X}|(1+3|w_{X}|)\\
d_{C^{1}}(6,1)=3(1-\Omega_{\mathrm{m},0}^{*})(1-A_{S})(1+3A_{S})\\
d_{C^{1}}(7,1)=3(1-\Omega_{\mathrm{m},0}^{*})(1-A_{S})(1+3A_{S}\alpha)\\
d_{C^{1}}(8a,1)=3(1-\Omega_{\mathrm{m},0}^{*})|w_{0}+1| \left \{ 1+ \left | \frac {3w_{0}(w_{0}+1)+w_{1}}{w_{0}+1} \right | \right \}\\
d_{C^{1}}(8b,1)=d_{C^{1}}(8a,1)\\
d_{C^{1}}(9,1)=8(1-\Omega_{\mathrm{m},0}^{*}-\Omega_{\Lambda,0})\\
d_{C^{1}}(10,1)=3(1-\Omega_{\mathrm{m},0}^{*})(1-A_{S})(1+3A_{S}|m+\frac{1}{2}|)\\
\end{array}
$$
In Table 3 we presented values of $d_{C^{0}}$ and $d_{C^{0}}$ for models from Table 1. The values of unknown parameters were estimating from SNIa data (Gold sample \cite{Riess:1999}) with fixed value of $\Omega_{\mathrm{m},0}^{*}=0.3$. Distances from the $\Lambda$CDM model are also illustrated on Figures 1A and 1B, respectively.
\begin{table}
\centering
\caption{Values of $d_{C^{0}}$ and $d_{C^{1}}$ for flat models from Table 1}
\begin{tabular}{|l|l|l|l}
\cline{1-3}
Case & \multicolumn{1}{c|}{$d_{C^{0}}$} & $d_{C^{1}}$ &  \\ 
\cline{1-3}
\multicolumn{1}{|c|}{2} & \multicolumn{1}{r|}{0.00} & \multicolumn{1}{r|}{0.00} &  \\ 
\cline{1-3}
\multicolumn{1}{|c|}{3} & \multicolumn{1}{r|}{0.70} & \multicolumn{1}{r|}{2.10} &  \\ 
\cline{1-3}
\multicolumn{1}{|c|}{4} & \multicolumn{1}{r|}{0.70} & \multicolumn{1}{r|}{3.50} &  \\ 
\cline{1-3}
\multicolumn{1}{|c|}{5} & \multicolumn{1}{r|}{0.60} & \multicolumn{1}{r|}{2.84} &  \\ 
\cline{1-3}
\multicolumn{1}{|c|}{6} & \multicolumn{1}{r|}{0.02} & \multicolumn{1}{r|}{0.08} &  \\ 
\cline{1-3}
\multicolumn{1}{|c|}{7} & \multicolumn{1}{r|}{0.02} & \multicolumn{1}{r|}{0.02} &  \\ 
\cline{1-3}
\multicolumn{1}{|c|}{8a} & \multicolumn{1}{r|}{1.34} & \multicolumn{1}{r|}{12.16} &  \\ 
\cline{1-3}
\multicolumn{1}{|c|}{8b} & \multicolumn{1}{r|}{1.22} & \multicolumn{1}{r|}{13.29} &  \\ 
\cline{1-3}
\multicolumn{1}{|c|}{9} & \multicolumn{1}{r|}{0.00} & \multicolumn{1}{r|}{0.00} &  \\ 
\cline{1-3}
\multicolumn{1}{|c|}{10} & \multicolumn{1}{r|}{0.02} & \multicolumn{1}{r|}{0.08} &  \\ 
\cline{1-3}
\end{tabular}
\end{table}
\begin{figure}[h]
\includegraphics[height=.2\textheight]{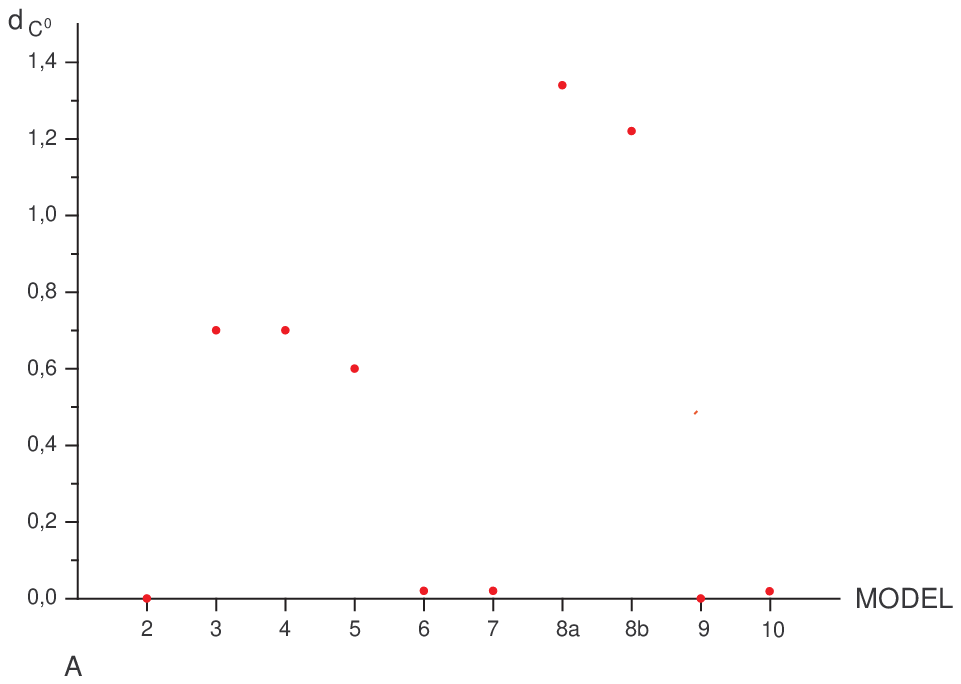}
\includegraphics[height=.2\textheight]{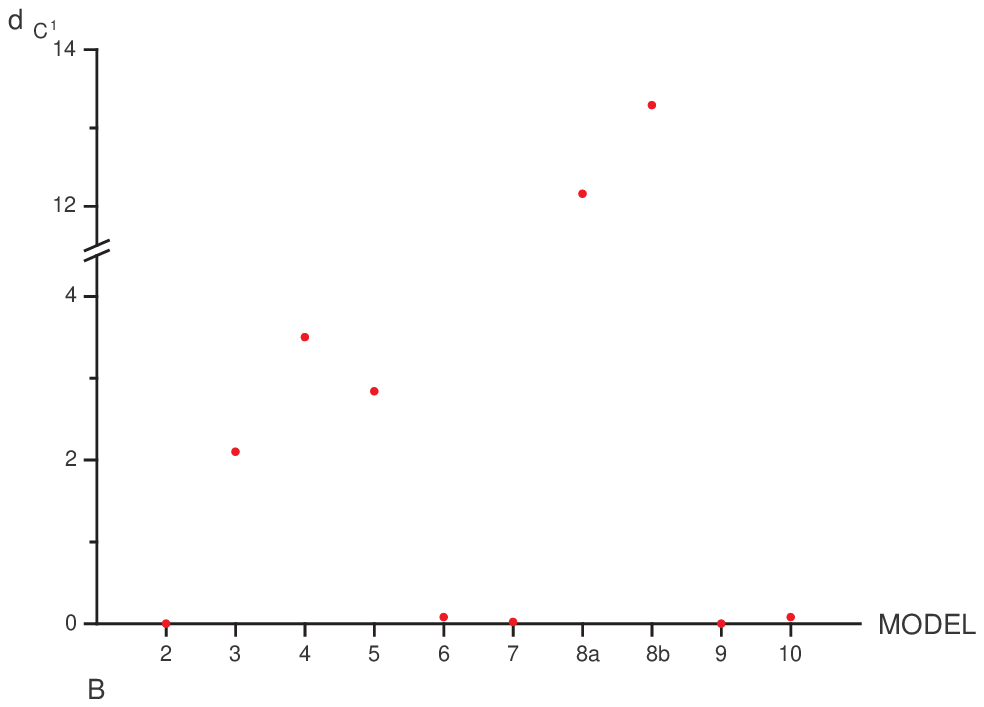}
\caption{A. $d_{C^{0}}$ and B. $d_{C^{1}}$ for flat models from Table 1}
\end{figure}

\section{Conclusion}

In the modern observational cosmology appeared the so-called degeneration problem manifested by the fact that many cosmological models which predict different evolution scenario in the past and in the future becomes still in good agreements with observational data. In the classical approach to model selection problem conclusion are made due to the best fit likelihood function for models under consideration, i.e. so called likelihood ratio test. Unfortunately this method allows to compare only nested models therefore it is not appropriate to compare cosmological models. In this context the Bayesian framework of model selection may be useful in decision which cosmological models is supported by observational data.

Recent observational data suggest that the coefficient of E.Q.S is very close to constant value $w \simeq -1$ but only $w=-1$ imply that energy density of dark matter (vacuum) is constant.

In this paper we characterize class of candidates models with dark energy in tools of Sobolev metric in the space of all dynamical system on the plane.

We believe that adequate model with dark energy should lies within the open ball at the centre in the $\Lambda$CDM model and with the small radius $\epsilon$.

Our convince based on the observation that the $\Lambda$CDM model is structurally stable, i.e. small changes r.h.s of the system (potential) does not disturb qualitative evolutional scenario of the model. Following the Peixoto theorem structurally stable dynamical systems on the plane form an open and dense subset in the space of all dynamical systems. Therefore structurally stable systems are typical (or generic) (structurally unstable becomes exceptional one) and can be approximated by adding small perturbations.

The metric introduced in the Banach space measure how so far we are from the $\Lambda$CDM - concordance model favored by the present observational data. With the help of this metric we can compare different proposition in a similar way like with the help of the Kullback-Leibler metric \cite{Burnham:2004}. 

Our main conclusion is that all simple dark energy models can be successfully unified within a scheme of dynamical systems of Newtonian type and potential function uniquely characterize different dark energy models. Some of them are generic like $\Lambda$CDM model and another exceptional. The models within close neighborhood of the $\Lambda$CDM model are typical because its structurally stable.

\end{document}